\documentclass{iopjournal}

\usepackage{graphicx}% Include figure files
\usepackage{dcolumn}% Align table columns on decimal point
\usepackage{bm}% bold math
\usepackage{braket}
\usepackage{hyperref}% add hypertext capabilities
\usepackage{color}
\usepackage{xcolor}
\usepackage{booktabs} 
\usepackage{amsmath,amssymb}
\usepackage[english]{babel}
\usepackage{microtype}
\usepackage{cite}
\begin{document}
\fancyhead[L]{}
%\articletype{Article type}

\title{Residual-Guided Global-Drive Design in Trapped-Ion Quantum Simulators}

\author{}
\author{{Yuanjing Zhang}, {Zewen Zhang$^{*}$}}

\affil{Shanghai Qi Zhi Institute, Shanghai 200232, PR China}

\email{zhangzewen@sqz.ac.cn}

\date{\today}% It is always \today, today,

%\keywords{ trapped ion, quantum simulation, Hamiltonian simulation}

\begin{abstract}
Trapped-ion quantum simulators are powerful platforms for exploring many-body physics, yet engineering of arbitrary interaction graphs remains a central challenge. To leverage the intrinsic structure of native motional modes and thereby enhance programmability,  a residual-guided optimization protocol is introduced for the systematic design of multi-tone global drive spectra. While the achievable interaction fidelity is fundamentally bounded by the available motional modes, this adaptive strategy converges on the theoretical limit using a number of laser tones that scales nearly linearly with system size. The versatility of this method enables the synthesis of both long-range and traditionally difficult short-range coupling topologies. Additionally, by incorporating a penalty for residual motional excitation, the algorithm effectively mitigates errors associated with incomplete closure of phase-space trajectories. Overall, this scalable framework significantly expands the programmable capabilities of trapped-ion architectures while maintaining low computational overhead and high potential for experimental implementation.
\end{abstract}


\section{Introduction}

Analog quantum simulators using trapped ions provide a flexible platform for studying quantum many-body physics in condensed matter~\cite{zhang2017observation,gorman2018engineering,manovitz2020quantum,kyprianidis2021observation,wang2024realizing,davoudi2020towards,richerme2023quantum,valahu2023direct,so2025quantum,foss2025progress} and high-energy physics~\cite{tan2021domain,schuster2022many,cai2022observation,jiang2022quantum}. Integrated with a small number of digital operations, their native dynamics extends to a wide range of accessible quantum evolutions~\cite{martinez2016compiling,maslov2018use,groenland2020signal,bravyi2022constant,richerme2025multi}.
The advantage of trapped ions in quantum simulation lies in their long-range interactions from Coulomb-mediated phonon modes, which thereby relax the connectivity constraints in many other architectures~\cite{blatt2012quantum}. Experiments using a global bichromatic M{\o}lmer-S{\o}rensen drive have demonstrated coherent simulation of Ising-type Hamiltonians with tunable long-range couplings~\cite{kim2010quantum, britton2012engineered}, revealing important dynamical features such as Kibble-Zurek mechanism~\cite{li2023probing}, emergent hydrodynamics~\cite{schuckert2020nonlocal,joshi2022observing}, squeezing~\cite{Bohnet2016,franke2023quantum}, and unconventional thermalization~\cite{neyenhuis2017observation,morong2021,kranzl2023experimental}. However, even with $\mathcal{O}(N)$ vibrational modes, trapped-ion simulation with a single global drive generally lacks tunability of interaction patterns.

Individually addressed schemes produce more general two-body interaction graphs~\cite{korenblit2012quantum,teoh2020machine,monroe2021programmable} and have been demonstrated in small-scale settings, including parallelized gates 
~\cite{figgatt2019parallel,grzesiak2020efficient,zhu2023pairwise,diaz2026arbitrary} and quantum simulations~\cite{mielenz2016arrays,lu2025implementing}. However, the control resources required typically grow quadratically with the number of ions~\cite{lu2019global,shapira2020theory}, making this approach difficult to scale up. Individual addressing is also less natural in Penning-trap settings~\cite{britton2012engineered,bohnet2016quantum,carter2026localized}, where global control is often preferred. 
A promising route to overcoming these constraints is to retain global addressing while increasing the number of bichromatic drives. Recent theoretical and experimental work has shown that stroboscopic~\cite{shapira2025programmable} and simultaneous~\cite{kyprianidis2024interaction} multi-tone global drives can realize high-accuracy power-law and sparse interaction graphs~\cite{peleg2026fast,huang2026large}. With additional single-qubit operations, multi-tone global-drive constructions can also support universal control in principle~\cite{richerme2025multi}. These results suggest that multiple global drives can improve Hamiltonian programmability without returning to the full overhead of individual addressing.

Designing a compact set of global drives for a target Hamiltonian remains challenging. The interaction depends nonlinearly on the beatnote frequencies and Rabi amplitudes, and the accessible interaction space is constrained by the trap normal modes~\cite{kyprianidis2024interaction}. Determining the optimal set of bichromatic tones required to approximate a natively inaccessible Hamiltonian to an allowed accuracy is mathematically a hard task. Practical issues further constrain the design, such as the limit of laser power,  the narrow  mode spacing at large scale, and residual spin-motion coupling. These constraints call for a resource-aware Hamiltonian synthesis method that gives a high-quality solution of a composite analog Hamiltonian while being efficient in both hardware and optimization overhead. In other words, we want an efficient algorithm to pick as few bichromatic beams as possible that approximate the target Hamiltonian to a satisfying accuracy.

This work introduces an  optimization method for constructing global drive designs in interaction graph engineering. The optimization incrementally adds global tones, guided by the current residual between the synthesized and the target interactions. The method primarily operates in the far-detuned regime, and it incurs little additional computational cost to incorporate a penalty on motional excitation, which efficiently mitigates errors arising from incomplete closure of phase-space trajectories. Through numerical demonstrations on a variety of representative long-range and short-range interaction targets, the algorithm consistently approaches the theoretical lower bounds of infidelity using a number of laser tones proportional to the system size. This reliable performance is maintained when extending to 2D ion crystals.

Our paper is structured as follows. Sec.~\ref{sec:model} introduces the multi-tone driven Hamiltonian, and presents the residual-guided drive design  procedure. Sec.~\ref{sec:results} demonstrates the application of global drive design on typical interaction patterns, and Sec.~\ref{sec:2d_prop}  extends to 2D implementation.
Sec.~\ref{sec:conclusion} concludes.

\section{The model}
\label{sec:model}
\subsection{Hamiltonian composition}
In ion traps, the native interaction under the M{\o}lmer-S{\o}rensen  protocol is approximated by long-range Ising models, 
\begin{equation}
    H = \sum_{i< j}J_{ij}\sigma_z^i\sigma_z^j,
\end{equation}
where $\sigma_z^i$ is the Pauli-$z$ matrix for the $i^\text{th}$ ion and $J_{ij}$ is the interaction strength between ions $i$ and $j$. 
For $N$ ions in the trap, a single pair of oppositely detuned lasers (bichromatic tone) labeled by $m$ generates the interaction matrix
\begin{equation}
    J_{ij}(\Omega_m,\mu_m) = \Omega_m^2 R \sum_{k=1}^N \frac{B_{ik} B_{jk}}{\mu_m^2 - \omega_k^2},
\end{equation}
where $R$ is the recoil energy, $\Omega_m$ is the effective Rabi frequency, $\mu_m$ is the beatnote frequency of the drive, and $B$ is the matrix of normal modes. Without loss of generality, we consider the highest $N$ vibrational modes with frequencies \{$\omega_k$\}, unless otherwise specified. $\{\Omega_m\}$ and $\{\mu_m\}$ are parameters to optimize in this work, and $\Omega_m$ is assumed to be uniform across all ions.

With $M$ bichromatic tones simultaneously on, the effective interactions add up to
\begin{equation}
    \widetilde J_{ij}=\sum_{m=1}^{M} J_{ij}(\Omega_m,\mu_m)
    =R\sum_{m=1}^{M}\Omega_m^2\sum_{k=1}^{N}\frac{B_{ik}B_{jk}}{\mu_m^2-\omega_k^2},
    \label{eq:multi_tone_interaction}
\end{equation}
where the summation over $m$ is the effect of all drives. Eq.~\eqref{eq:multi_tone_interaction} assumes the far-detuned regime in which residual spin-motion entanglement is perturbative and largely suppressed.

In this work, the search space is restricted to $2N$ tones, formed by placing a pair of red- and blue-detuned tones in the vicinity of each vibrational mode. This candidate set of tones is chosen for convenience and is sufficiently large for the optimization to reach the theoretical limit, but the search space can certainly be expanded if necessary. The ultimate task of this work is to select the minimal number of tones out of the pool and to assign the Rabi and the beatnote frequencies of these tones, so that the synthesized interactions can approximate target interaction patterns to a certain accuracy. 
Accordingly, 
\begin{equation}
\label{eq:composite}\widetilde{J}_{ij}^\text{opt}=\sum_{m=1}^{2N}\chi_m J_{ij}(\Omega_m,\mu_m)
    =R \sum_{m=1}^{2N}\chi_m\Omega_m^2\sum_{k=1}^{N}\frac{B_{ik}B_{jk}}{\mu_m^2-\omega_k^2},
\end{equation}
where the binary variable $\chi_m\in\{0,1\}$ is introduced to mark whether tone $m$ is on ($1$) or off ($0$), and we optimize only the parameters of the tones in use. 
Recent works~\cite{shapira2025programmable,kyprianidis2024interaction} proposed a scheme of using at least $2N+1$ equal-spacing tones for Hamiltonian synthesis, which automatically decouples the ion motion from the spin degree of freedom. However, for large systems, this scheme requires complex optical alignment. Aiming to reduce hardware complexity and achieve better pragmatic performance than these protocols, this study tackles the challenge of efficiently optimizing a hybrid parameter space comprising integer and real variables 
\{$\chi_m,\mu_m,\Omega_m$\}, such that the optimized quantity $\widetilde{J}^\text{opt}$ agrees closely with the desired interaction profile $\widetilde J^{\rm tar}$. 

To simplify the notation,  $\mathcal J^{(k)}_{ij}= B_{ik}B_{jk}$ is introduced to record the $k^\text{th}$ mode contribution matrix and the synthesized interaction matrix~\eqref{eq:composite} is rewritten as
\begin{equation}
    \widetilde J^\text{opt}_{ij}=\sum_{k=1}^{N} c_k \mathcal J^{(k)}_{ij},
    \qquad c_k=R\sum_m^{2N}\frac{\chi_m\Omega_m^2}{\mu_m^2-\omega_k^2}.
\end{equation}
Thus, the multi-tone design problem is decomposed into two steps: determining the coefficients $\{c_k\}$ and solving for \{$\chi_m,\mu_m,\Omega_m$\} given the set of  $\{c_k\}$. 
The accessible interaction patterns  lie in the mode-space span~\cite{kyprianidis2024interaction}
\begin{equation}
    \mathcal S=\mathrm{span}\{\mathcal J^{(1)},\ldots,\mathcal J^{(N)}\}
\end{equation}
with $N-1$ degrees of freedom. As an arbitrary interaction pattern has $N(N-1)/2$ independent entries, in many cases, it cannot be perfectly realized with multi-tone composition and there is a theoretical upper limit on the fidelity. Given a target $\widetilde J^{\rm tar}$, the best approximate interaction allowed within the mode space is its projection on $\mathcal S$
\begin{equation}
    \widetilde J^\star=\Pi_{\mathcal S}\widetilde J^{\rm tar}.
\end{equation}
The algorithm in this work helps to find a small number of experimentally feasible global tones to approach $\widetilde J^\star$.

\begin{figure}[h]%
\centering
\includegraphics[width=0.95\linewidth]{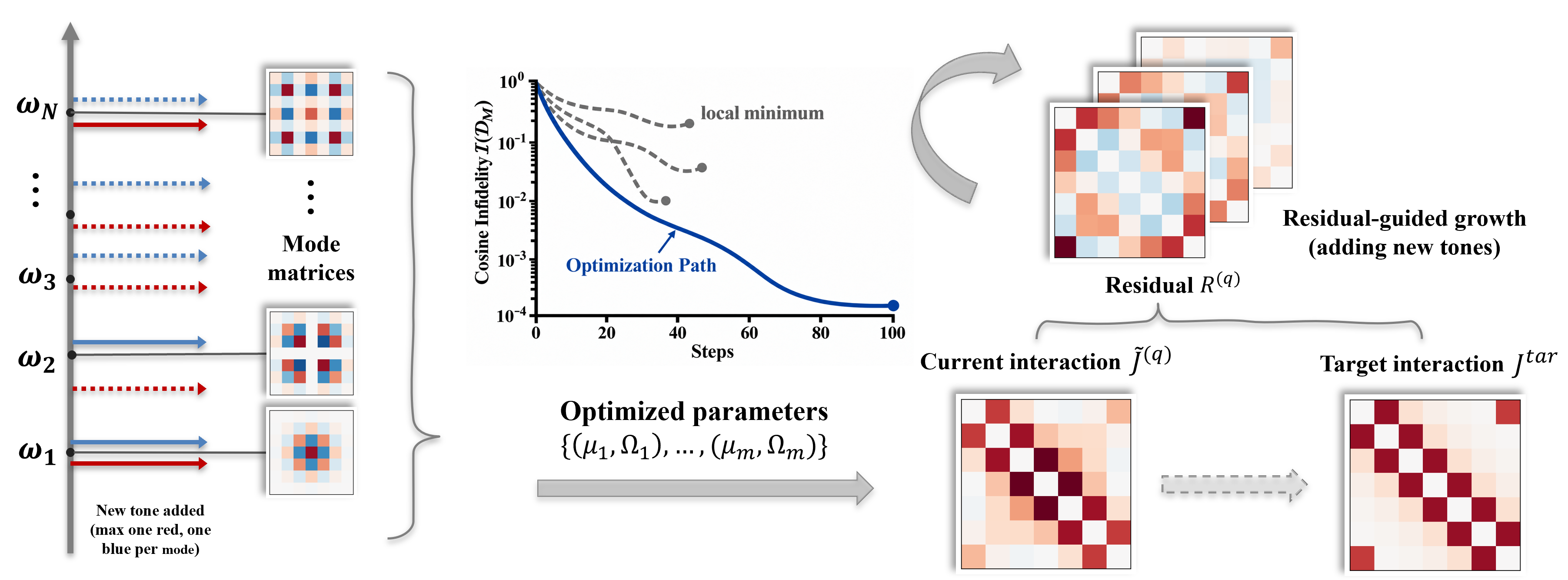}
\caption{\label{fig:1} {Schematic workflow of global-drive Hamiltonian design using the residual-guided protocol. The left panel illustrates the procedure of selecting a subset of candidate tones from the mode spectrum, where solid arrows denote the chosen tones and dashed arrows mark the remaining tones in the pool. The selected tones are then used to synthesize the target Hamiltonian, with their $\mu_m$ and $\Omega_m$ optimized toward the lowest infidelity achievable at the current step. As indicated in the right panel, the number of chosen tones increases along a residual-guided track and eventually reaches the tolerated infidelity. Although each tone's detuning is constrained to remain sufficiently large, a closure penalty can be naturally included  to suppress motional excitations.}}
\end{figure}

\subsection{Optimization}

To tackle the multi-tone design problem, this work introduces a residual-guided strategy, which is widely implemented in  different areas of learning tasks~\cite{2017DnCNN,Ke2017}. This strategy avoids the exponential complexity of exhaustively enumerating the combinations of candidate tones.
The residual-guided drive design (RGDD) is based on a simple observation: for many interesting interaction patterns, only a few tones are needed to approximate the patterns to  a given accuracy; however, assigning the frequencies of chosen tones remains challenging. Rather than jointly optimizing all tones, RGDD constructs the drive set incrementally, reducing the dimensionality of optimization. At each step, the current residual marks the part of the target graph that remains poorly represented, and the next global tone is chosen to maximally reduce this residual. 

Next to each mode $\omega_k$, the pairs of red- and blue-detuned candidate tones make up a pool of beatnote frequencies  
\begin{equation}\mu^\pm_{m_k}=\omega_k\pm\delta^\pm_{m_k}\Delta_{m_k},
\end{equation}
where $\delta^\pm_{m_k}$ sets the detuning range to optimize and $\Delta_{m_k}$ is the spacing between neighboring modes, \textit{i.e.}, $\omega_k-\omega_{k-1}$ for red-detuned candidates and $\omega_{k+1}-\omega_k$ for blue-detuned, and the highest and lowest candidates are set properly.  $\delta^\pm_{m_k}$ are constrained to $[0.01,0.5)$ in optimization, with the lower limit controls heating of the phonon modes. During the optimization, we introduce a logarithmic variable $r_m=\log \Omega_m$ to enforce that the Rabi frequencies remain positive. 

The objective takes a scale-invariant format as cosine infidelity~\cite{shapira2020theory},
\begin{equation}
    \mathcal I(\mathcal D_M)=\frac{1}{2}\left(1-\frac{\left\langle \widetilde J(\mathcal D_M),\widetilde J^{\rm tar}\right\rangle_F}{\|\widetilde J(\mathcal D_M)\|_F\,\|\widetilde J^{\rm tar}\|_F}\right),
    \label{eq:cosine_infid}
\end{equation}
where $\langle.,.\rangle_F$ and $\| .\|_F$ denote the Frobenius inner product and Frobenius norm, respectively, and $\mathcal D_M$ is the set of optimized laser parameters for the selected tone set $M$. Infidelity depends only on the angle between the synthesized and target interaction matrices in the $\tilde J$-space.

The ideal Ising dynamics generated by $\widetilde J(\mathcal D_M)$  requires that the spins decoupled from ion motion at the end of the evolution, \textit{i.e.}, the phase-space trajectory of each mode returns to its origin at the gate time $t_g$. Although we impose a lower limit on the detuning, the displacement of modes still accumulates over time. With $M$ bichromatic tones, mode $k$ retains a residual displacement,
\begin{equation}
    \alpha_k(t_g)=\eta_k\sum_{m=1}^M\frac{\Omega_m}{2}\,
    \frac{1-e^{i(\mu_m-\omega_k)t_g}}{\mu_m-\omega_k},
    \qquad \eta_k=\sqrt{\frac{R}{\omega_k}},
    \label{eq:alpha_k}
\end{equation}
where $\eta_k$ is the Lamb-Dicke parameter of mode $k$. We conservatively take $R$ to be $2\pi \times 50\;\mathrm{kHz}$, which should be an order of magnitude higher than its values in real experiments. To the first order, the total motional fidelity penalty~\cite{roos2008ion} is estimated by the quantity
\begin{equation}
    A_\alpha=\sum_{k=1}^{N}\left|\alpha_k(t_g)\right|^2.
    \label{eq:A_alpha}
\end{equation}
To maintain the fidelity of $\widetilde J(\mathcal D_M)$, the term $A_\alpha$ must be kept low. Optimizing $\mathcal I$ alone does not guarantee suppression of $A_\alpha$. Therefore, the design objective incorporates a closure penalty,
\begin{equation}
    \mathcal L(\mathcal D_M)=\mathcal I(\mathcal D_M)
+\lambda_\alpha\left[\max\!\left(0,\,\ln\frac{A_\alpha}{A_{\rm tar}}\right)\right]^2,
    \label{eq:closure_objective}
\end{equation}
where $\lambda_\alpha$ is the penalty weight and $A_{\rm tar}$ sets the closure error tolerance. The penalty vanishes for closure-compliant drives ($A_\alpha\leq A_{\rm tar}$) and grows quadratically in $\ln(A_\alpha/A_{\rm tar})$ otherwise. $A_{\rm tar}$ is set to 1 in this work. The objective remains continuous at $A_\alpha=A_{\rm tar}$ and is therefore compatible with gradient-based optimizers. 
By scanning $\lambda_\alpha$, we choose small $\lambda_\alpha=0.003$--$0.03$, which has little impact on the infidelity minimization while reducing the closure error by two orders of magnitude relative to the unpenalized optimization.

Although a large $t_g$ clearly eliminates $A_\alpha$, we keep $t_g$ at a finite value to maintain practical efficiency. Given $\bar{\Delta}$ as the mean mode spacing, the typical gate time is chosen as~\cite{peleg2026fast} $t_{g_0}=2\pi/0.1\bar{\Delta}$. 
For optimization including the closure error term, we scan $t_g=s\,t_{g_0}$ with $s$ ranging from $0.5$ to $200$ and choose the smallest $t_g$ satisfying $A_\alpha \leq A_{\rm tar}$ for all $N$ with given targets and $M$. During the scan (see Appendix~\ref{app:closure_tradeoff}), the target interaction scale is kept as $J_0t_g=\tfrac{\pi}{4}$, where $J_0$ is the maximum coupling strength of the target interaction, so that all the target Ising evolutions are scaled identically.

%############################

RGDD consists of two stages: initialization and residual-guided growth. The optimization is initialized by ranking each mode's contribution  to the target interaction matrix, which is the projection coefficients $c_k^{\rm tar}=\langle  {\mathcal J}^{(k)},\widetilde J^{\rm tar}\rangle$. Accordingly, we select a small set of $k_0$ modes with the largest $|c_k^{\rm tar}|$. For each of these $k_0$ modes, we greedily assign either a red- or blue-detuned tone in its vicinity  by minimizing the infidelity~\eqref{eq:cosine_infid}. 
Naturally, this step warm-starts the optimization with a set of tones that potentially carry the major contributions to the target interaction pattern. 

After initialization, RGDD incrementally adds tones through a residual-guided growth process. With  $q$ tones incorporated into the model and their laser parameters $\mathcal D_q$ optimized via gradient descent, the resulting synthesized interaction $\widetilde J^{(q)} =\widetilde J(\mathcal D_q)$ gives a residual 
\begin{equation}
\label{eq:residual}
R^{(q)} = \widetilde{J}^{\rm tar} - \widetilde{J}^{(q)},
\end{equation}
which represents the deviation of the synthesized interaction matrix from the target and efficiently guides the placement of the next tone. 
The contribution of different modes to reducing the residual is measured by the coefficient $c_k^{(q)}= \langle R^{(q)}, {\mathcal J}^{(k)}\rangle$. A subset of $k_q$ modes with the largest $|c_k^{(q)}|$ serves as the pool for a new tone to add.
During the optimization, each candidate in the pool of $2N$ tones can be added only once.
With the optimized parameters $\mathcal D_q$ for the previous $q$ tones frozen, the tone that gives the lowest infidelity $\mathcal I$ among the $k_q$ candidates is selected.
After the $(q+1)^\text{th}$ tone is selected, a joint refinement is performed to get the optimal parameter set $\mathcal D_{q+1}$. 
To avoid trapping in local minima, perturbations are repeatedly added to the previous $\mathcal D_{q}$ before gradient descent, and the best $\mathcal D_{q+1}$ is adopted from 10 random seeds per data point, selected by the lowest infidelity.
If the refined infidelity still exceeds the requirement, the algorithm proceeds to another round of residual-guided growth, with infidelity decreasing monotonically as tones are added.
The algorithm in this work uses the L-BFGS-B optimizer.

\section{Results on 1D Ion Crystal}
\label{sec:results}

Through trials on multiple important target interaction patterns, the results in this section demonstrate that RGDD facilitates the construction of high-fidelity Hamiltonians with roughly $O(N)$ tones, and nearly guarantees reaching the theoretical limit.
The lower bound of infidelity $\mathcal{I}_{\mathrm{lb}}$ is obtained by an unconstrained least-squares projection onto the mode space~\cite{kyprianidis2024interaction},
which gives the set of  mode coefficients $\{c_k^{\rm tar}=\langle  {\mathcal J}^{(k)},\widetilde J^{\rm tar}\rangle\}$ for all $N$ modes. 
To benchmark the performance of RGDD,  a baseline with  fixed-frequency tones is used for comparison. The baseline employs a nonnegative least-square (NNLS) construction to get the optimal infidelity with a preset number $M$ of drive tones. The tones are placed near the modes with the largest $|c_k^{\rm tar}|$ and their frequencies $\{\mu_m\}$ are preset.
The Rabi frequencies $\Omega_m$ in NNLS are optimized to close the gap between the multiple-tone design and the optimal set of $\{c_k^{\rm tar}\}$, \textit{i.e.}, $\min_{\Omega_m\geq 0}\|\sum_m\Omega_m^2R/(\mu_m^2-\omega_k^2)-c^{\rm tar}_k\|$. 
Since NNLS sometimes cannot reach the required fidelity with only $2N+1$ tones, in practice we allow it to use more than $2N+1$ tones. In addition, the comparison with equal-spacing tones outside the adiabatic regime is discussed.

\subsection{Power-law Interactions}
\label{sec:1d-powerlaw}
The first example of RGDD is its application to power-law interaction targets $J_{ij}^{\mathrm{tar}} \propto |i-j|^{-\alpha}$ in a Paul trap with axial frequency $\omega_z = 2\pi \times 0.1\;\mathrm{MHz}$ and transverse frequency $\omega_x = 2\pi \times 5\;\mathrm{MHz}$. Typical exponents are tested, including  long-ranged ($\alpha = 1.5$ and $2.5$) patterns that are commonly recognized as native to hardware, as well as short-ranged ($\alpha = 4.0$) patterns. For each $(N, \alpha)$, the number of drive tones $M$ is swept, and the minimum $M_\text{min}$ required to achieve target infidelity $\mathcal{I} \leq 0.05$ and $\mathcal{I} \leq 0.01$ is recorded.

\begin{figure}[h]
\centering
\includegraphics[width=\textwidth]{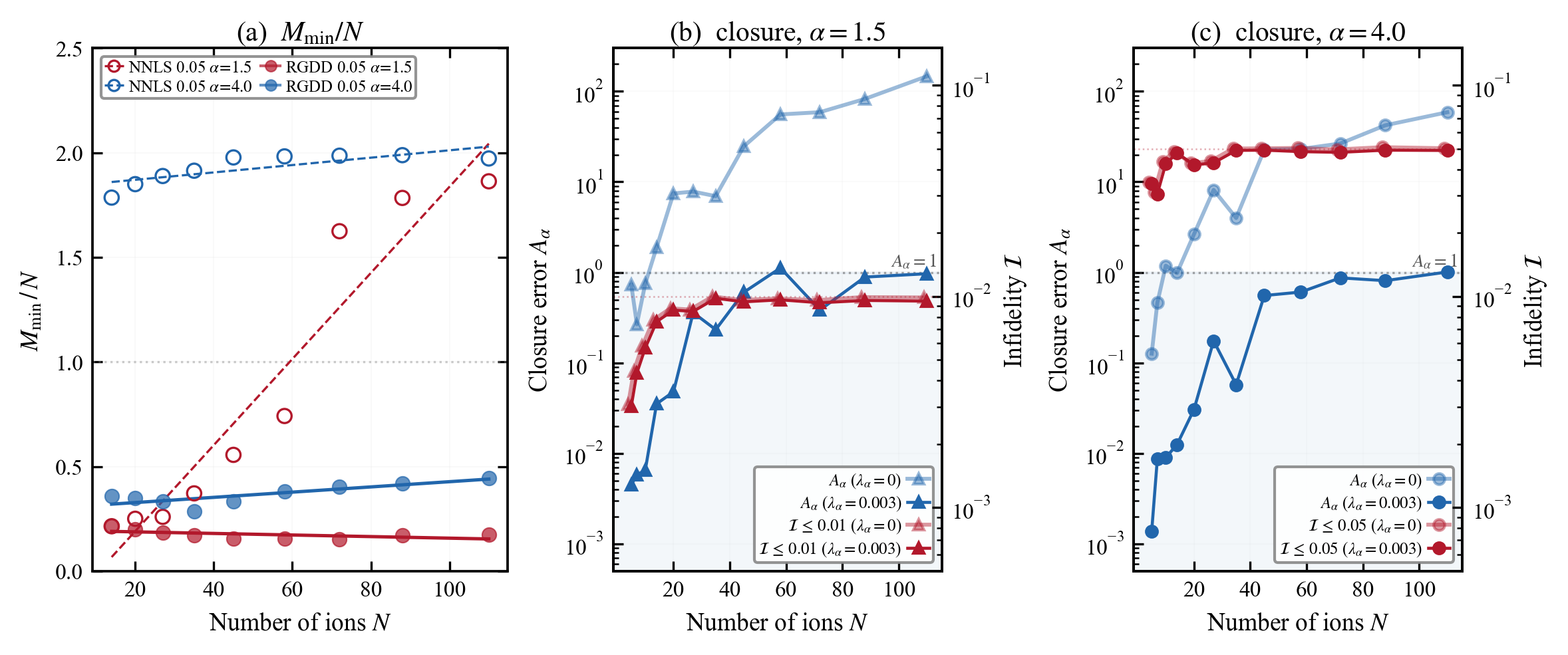}
\caption{Tone-count scaling, infidelity, and estimate of closure error for power-law targets. (a) Number of tones per ion $M_{\min}/N$ versus ion number $N$ for $\alpha = 1.5$ (red) and $\alpha = 4.0$ (blue) at the fidelity threshold  $\mathcal{I} \leq 0.05$. Open markers with dashed lines: fixed-frequency NNLS baseline; filled markers with solid lines: RGDD results. The lines are linear fits. 
(b) and (c) Closure error $A_\alpha$ (left axis, blue) and infidelity $\mathcal{I}$ (right axis, red) versus $N$ for $\alpha = 1.5$ and $\alpha = 4.0$, respectively. Light markers: unpenalized solutions ($\lambda_\alpha = 0$); dark markers: closure-aware optimization ($\lambda_\alpha = 0.003$). 
Infidelity curves in (b) and (c) are slightly offset for visibility.
}
\label{fig:M_scaling_bfgs}
\end{figure}

In these power-law examples, RGDD sharply outperforms the naive NNLS method. The scaling results for ion numbers $N = 14$--$110$ across two interaction ranges are summarized in Fig.~\ref{fig:M_scaling_bfgs}(a), which plots the number of tones per ion $M_{\min}/N$. For the long-range interaction example ($\alpha=1.5$), when comparing the average number of tones required to reach an infidelity of 0.05, the RGDD results exhibit a decreasing trend with increasing ion number, whereas the NNLS results show the opposite, increasing trend. This observation supports the conclusion that RGDD can serve as a highly reliable tool for composing long-range interactions. For the short-range interaction example ($\alpha=4.0$), although both methods exhibit slowly increasing trends, the RGDD results clearly prevail by a large prefactor.

To quantitatively characterize the performance of the optimization algorithms, we fit the number of tones per ion with the system size $N$ according to $M_{\min}/N = pN + q$.
The resulting slope $p$ serves as a measure of the algorithm's scaling behavior.
For $\mathcal{I} \leq 0.05$, RGDD maintains a flat scaling across all tested power-law exponents, while NNLS shows a growing $M_{\min}/N$ for the long-range interactions. Specifically, RGDD gives $p=-0.4$, $-1.2$, and $1.2$ for $\alpha=1.5$, $2.5$, and $4.0$, respectively, while NNLS gives $p=20.6$, $9.4$, and $1.8$ in the same order.  Under the tighter threshold $\mathcal{I} \leq 0.01$, RGDD achieves $M_{\min}/N = 0.68$ at $N = 110$ for $\alpha = 1.5$, compared with approximately 2 for NNLS. For $\alpha = 4.0$,  the theoretical lower bound at small system sizes is higher than $0.01$ for these shorter-range interactions at $N = 5$. At large size of $N=110$, RGDD gives $M_{\min}/N = 0.45$ for $\alpha = 4.0$. NNLS requires $M_{\min}/N = 1.97$ for the same settings.

% \begin{table}[h]
% \centering
% \caption{Fitted slopes $p$ (in units of $10^{-3}$) in $M_{\min}/N = pN + q$ for each method, fidelity threshold, and interaction exponent $\alpha$.}
% \label{tab:scaling_exponents}
% \begin{tabular}{ccccccc}
% \toprule
%  & \multicolumn{2}{c}{$\alpha = 1.5$}
%  & \multicolumn{2}{c}{$\alpha = 2.5$}
%  & \multicolumn{2}{c}{$\alpha = 4.0$} \\
% \cmidrule(lr){2-3}
% \cmidrule(lr){4-5}
% \cmidrule(lr){6-7}
% $\mathcal{I}$ & NNLS & RGDD & NNLS & RGDD & NNLS & RGDD \\
% \midrule
% $\leq 0.05$ & 20.6 & $-0.4$ & 9.4 & $-1.2$ & 1.8 & 1.2 \\
% $\leq 0.01$ & 1.0 & 3.1 & --- & --- & --- & --- \\
% \bottomrule
% \end{tabular}
% \end{table}

The inclusion of the closure penalty in the objective~\eqref{eq:closure_objective} serves to keep the accumulated displacement at a low level. Figs.~\ref{fig:M_scaling_bfgs}(b) and (c) present the phase-space closure error $A_\alpha$ for the optimization results at given $N$, and compare these results with those obtained  without penalty. 
For the case of $\alpha=1.5$ and a target infidelity $\mathcal{I}\leq 0.01$ [Fig.~\ref{fig:M_scaling_bfgs}(b)], the penalized objective with $\lambda_\alpha = 0.003$ produces solutions with $A_\alpha<1.0$ at the largest system size considered. By contrast, when the penalty is omitted ($\lambda_\alpha = 0$), the resulting closure errors are approximately two orders of magnitude larger than those obtained with the penalty.
These findings also apply to shorter-range interactions, as demonstrated by the $\alpha=4.0$ results in Fig.~\ref{fig:M_scaling_bfgs}(c).
The inclusion of the penalty term in the objective function proves essential during optimization to maintain  low excitations. Importantly, the results show that including the closure penalty in the objective at a certain level does not significantly affect the final optimized infidelity, which ensures the practical viability of the algorithm. Details on the joint role of $\lambda_\alpha$ and the gate time $t_g$ are provided in Appendix~\ref{app:closure_tradeoff}.

\subsection{Short-range models}
\label{sec:1d_short range}
Beyond the $\alpha=4.0$ examples, the demonstration of RGDD also extends to other short-range models, supporting the algorithm's robust performance. This section presents the optimization results for short-range interaction patterns, which include the long-tail Yukawa potential as well as nearest-neighbor models such as $J_1$-$J_2$, the Su--Schrieffer--Heeger (SSH), and dimerized chains. To improve optimization performance on short-range models, where distant interactions are identically zero, we introduce an edge-weighted residual guidance for candidate selection. Specifically, nearest-neighbor, next-nearest-neighbor, and other longer-range edges are assigned distinct weights, and the weighted residual is used together with the original residual to rank candidate modes for their importance. The final tone selection and parameter optimization are still evaluated by the original cosine infidelity. The detailed weighting scheme is provided in Appendix~\ref{app:edge_weight}.

\begin{figure}[h]
\centering
\includegraphics[width=\textwidth]{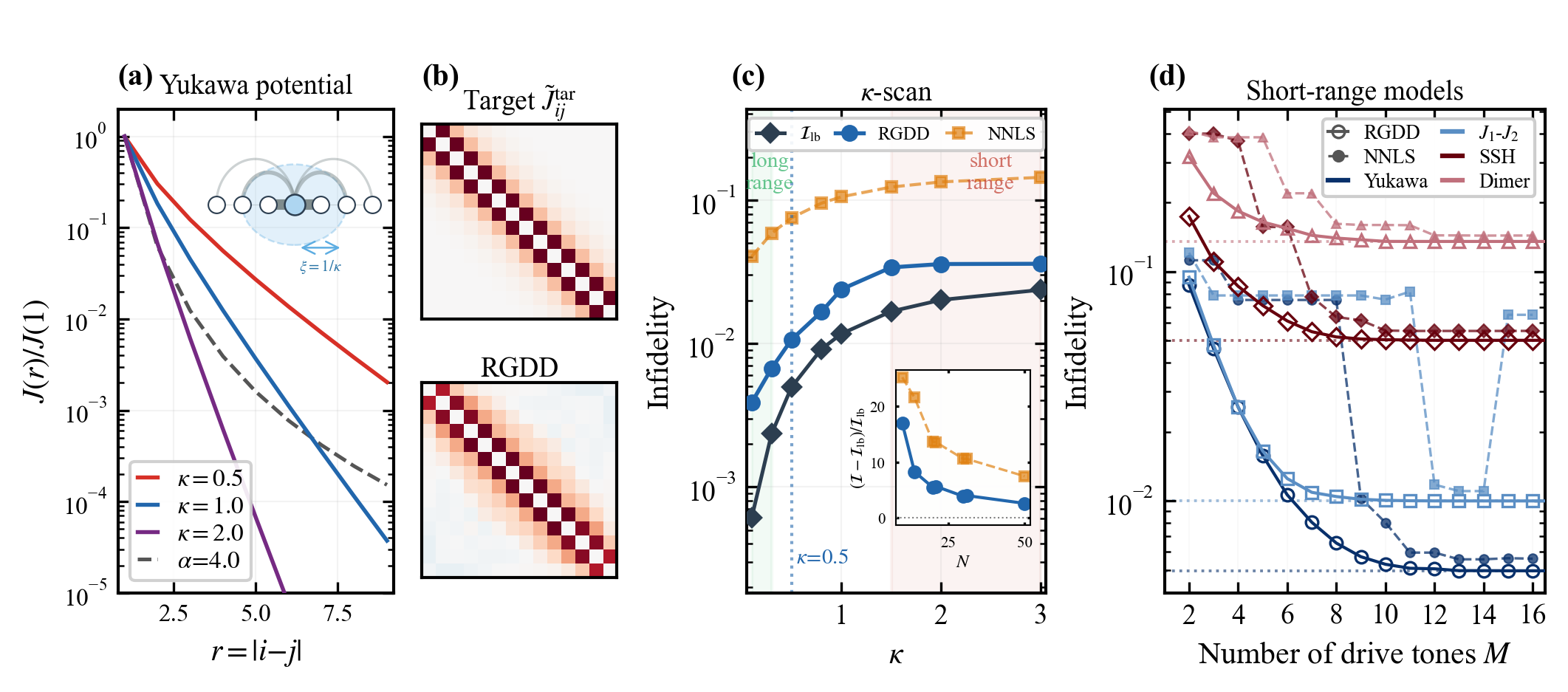}
\caption{(a)~Yukawa interaction decay $J(r)/J(1)$ for screening parameters $\kappa = 0.5$, $1.0$, and $2.0$, along with the power-law interaction with  $\alpha = 4.0$ for reference. Inset: schematic of Yukawa interactions with the screening cloud of extent $\xi = 1/\kappa$, where the bond width denotes the assigned weights. (b)~Target $\tilde{J}^{\mathrm{tar}}_{ij}$ and the RGDD-achieved interaction matrix for $\kappa = 0.5$ with 14 global drives. (c)~Infidelity versus $\kappa$ at fixed $M = 6$: theoretical lower bound $\mathcal{I}_{\mathrm{lb}}$ (diamonds), RGDD results (circles), and NNLS results (squares). Inset: relative gap $(\mathcal{I} - \mathcal{I}_{\mathrm{lb}})/\mathcal{I}_{\mathrm{lb}}$ versus ion number $N$ at $\kappa = 0.5$, $M \approx 0.2N$. (d)~Infidelity versus tone count $M$ for four short-range targets at $N = 14$. Open markers and solid lines denote the RGDD results, while filled markers and dashed lines denote the NNLS results. Horizontal dotted lines: theoretical lower bound $\mathcal{I}_{\mathrm{lb}}$.}
\label{fig:yukawa_demo}
\end{figure}

The Yukawa interaction, with its tunable range, serves as a versatile testbed for engineering interaction patterns in quantum simulators. Its functional form $J_{ij} = e^{-\kappa|i-j|}/|i-j|$ was originally proposed in the study of meson-mediated nuclear forces~\cite{Yukawa1935}. The  range of the interaction is set by a screening parameter $\kappa$: the coupling decays exponentially with distance beyond the screening length $\xi = 1/\kappa$. In the  $\kappa \to 0$ limit, the potential reduces to a long-range $1/r$ law. For large $\kappa$, the interaction is dominated by nearest-neighbor couplings.
Fig.~\ref{fig:yukawa_demo}(a) shows some representative patterns of the Yukawa potential at different values of $\kappa$.

Using only a small number of global tones, RGDD is capable of synthesizing the Yukawa interaction  with moderately high fidelity, even in the short-range regime corresponding to large $\kappa$. As an illustration, Fig.~\ref{fig:yukawa_demo}(b) provides a typical Yukawa interaction pattern and its RGDD-optimized approximation. Fig.~\ref{fig:yukawa_demo}(c) shows the RGDD-optimized infidelity at various $\kappa$ with fixed $M = 6$, which is compared with the theoretical lower bound $\mathcal{I}_{\mathrm{lb}}$ and the NNLS results. Across the full range of $\kappa$, the infidelity achieved by RGDD closely tracks the lower bound $\mathcal{I}_{\mathrm{lb}}$, deviating by only a small factor, whereas the NNLS results lie more than an order of magnitude above the bound. This performance gap persists across tone counts $M > 6$.  
The inset highlights the advantage of RGDD across system sizes: at $\kappa = 0.5$ with $M \approx 0.2N$, the relative gap $(\mathcal{I} - \mathcal{I}_{\mathrm{lb}})/\mathcal{I}_{\mathrm{lb}}$ of NNLS remains large from $N = 10$ to $50$, while that of RGDD falls steadily with system size and stays well below the NNLS values.
In addition, the results show that, with a small number of tones, long-range models can be better represented through this multiple global drive design, which reflects the long-range nature of interactions in trapped ions~\cite{islam2013emergence,richerme2014non}.

As shown in Fig.~\ref{fig:yukawa_demo}(d), the global drive design also extends to  three  other short-range interaction models: the $J_1$--$J_2$ chain with up to next-nearest-neighbor interactions, the  SSH model with alternating nearest-neighbor interactions, and a dimer covering. Throughout these examples, RGDD approaches the $\mathcal{I}_{\mathrm{lb}}$ using roughly $N/2$ tones, confirming the effectiveness of the optimization. Its stable advantage over NNLS highlights the critical role of including $\mu$ in the optimization procedure. The NNLS infidelity is non-monotonic in $M$, \textit{e.g.}, it rises by a factor of six for the $J_1$--$J_2$ chain from $M=14$ to $15$, because a tone placement fixed in advance can degrade as $M$ grows.  
In the Appendix~\ref{app:nnls_convergence}, we further compare RGDD against the non-adiabatic protocol~\cite{shapira2020theory,kyprianidis2024interaction}, which requires a minimum of ($2N+1$) equally spaced tones. However, in our trials, this protocol cannot reach the infidelities as low as those achieved by the RGDD optimization. Although increasing the tone count can enhance the performance of the equal-spacing method, it also raises the complexity of optical alignment.

\section{Extension to 2D Settings}
\label{sec:2d_prop}
 
The RGDD method extends naturally to two-dimensional (2D) ion crystals, which offer richer mode structures. In a Paul trap, a planar crystal can be reached by tuning the aspect ratio of the confining potential, which drives the ion chain through a zigzag transition as the transverse confinement is weakened relative to the axial confinement~\cite{kiesenhofer2023controlling,Guo2024}.  Penning trap instead achieves a planar ion crystal through static electric and magnetic fields together with a rotating wall potential~\cite{Bohnet2016,Britton2012}. Both routes can be scaled up to hundreds of ions. 
As the mode contribution retains the form $J_{ij}^{(k)} = B_{ik} B_{jk}$, RGDD applies here directly. In this section, only the out-of-plane (drumhead) modes are included in the model, a restriction justified by their spectral isolation from the remaining modes. Including in-plane modes in the optimization reduces the theoretical bound to infidelity by only a small factor, as quantified in Appendix~\ref{app:2d_mode_analysis}.

Fig.~\ref{fig:2d_combined} applies RGDD to a Paul-trap ion crystal with trap frequencies $\omega_x = 2\pi\times0.147\;\mathrm{MHz}$, $\omega_y = 2\pi\times0.623\;\mathrm{MHz}$, and $\omega_z = 2\pi\times2.200\;\mathrm{MHz}$, yielding a planar zigzag crystal structure for ion numbers $N = 14$--$42$. We present optimization results targeting a 2D power-law interaction $J_{ij}^{\mathrm{tar}} \propto |\mathbf{r}_i - \mathbf{r}_j|^{-1.5}$, where $\mathbf{r}_i$ are the 2D equilibrium positions of ion $i$. An example of the equilibrium positions and interaction pattern is shown in Fig.~\ref{fig:2d_combined}(a).  
Fig.~\ref{fig:2d_combined}(b) presents a comparison between the target interaction pattern and the RGDD-optimized pattern, whose fidelity is high, especially in the center of the ion crystal. Typical errors reside in the weak entries at the marginal positions.

\begin{figure}[h]
\centering
\includegraphics[width=\textwidth]{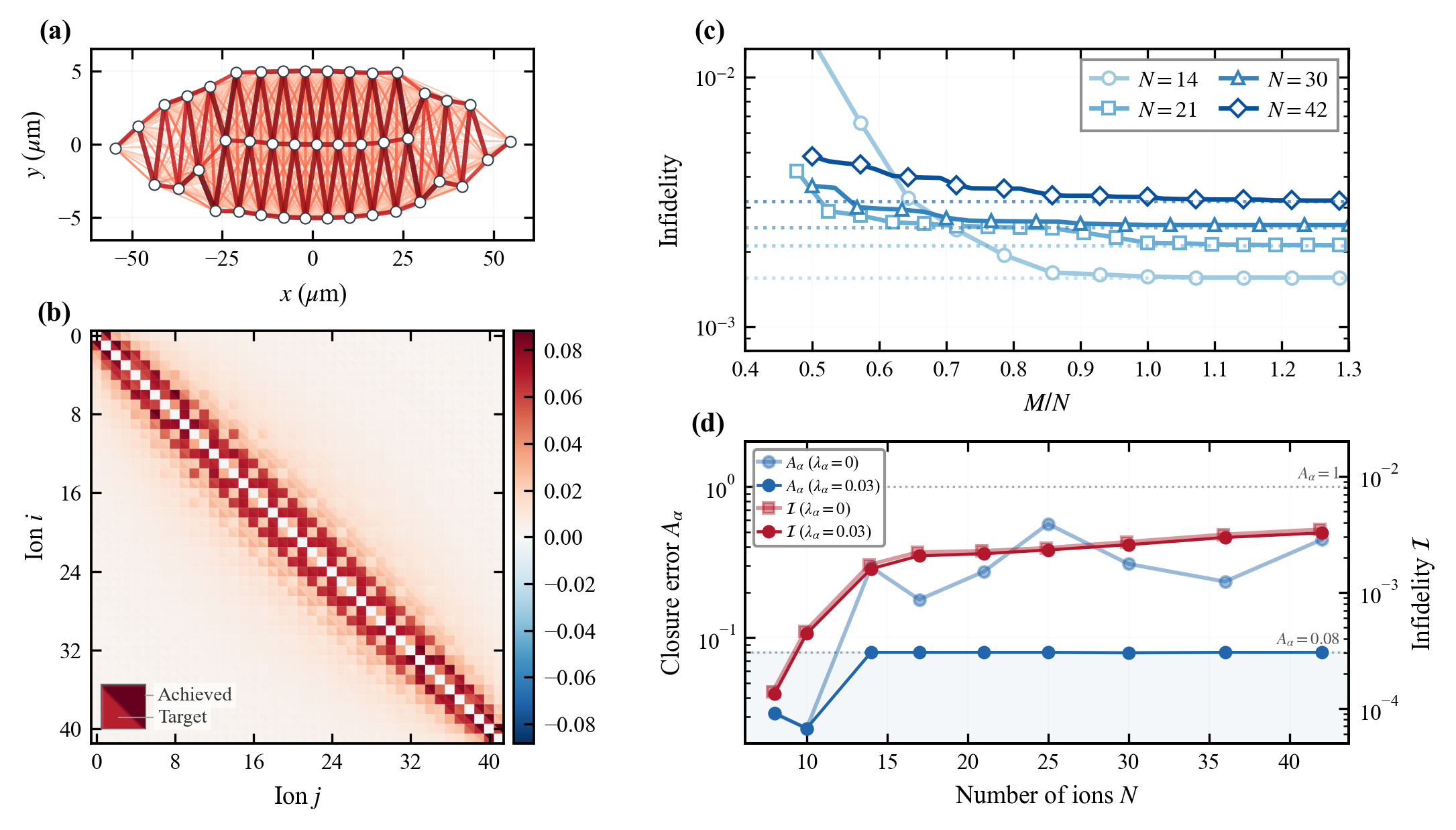}
\caption{ (a)~Target interaction $J_{ij}^{\mathrm{tar}} \propto |\mathbf{r}_i-\mathbf{r}_j|^{-1.5}$ visualized as bonds on the 2D ion crystal ($N = 42$). Brightness and thickness reflect the coupling strength. (b)~Split-cell interaction matrix for $N = 42$: the bottom-left triangles show the target $\tilde{J}^{\mathrm{tar}}_{ij}$ and the top-right triangles show the optimized $\tilde{J}_{ij}$ at the lowest infidelity achieved with RGDD. (c)~Infidelity $\mathcal{I}$ versus average tone count $M/N$ for various $N$. Horizontal dotted lines mark the theoretical lower bound $\mathcal{I}_{\mathrm{lb}}$.  (d)~Closure error $A_\alpha$ (left axis, blue) and infidelity $\mathcal{I}$ (right axis, red) versus $N$ at the optimized settings. Light markers:  solutions with $\lambda_\alpha = 0$. Dark markers: closure-aware optimization with $\lambda_\alpha = 0.03$. Infidelity datasets are slightly offset for visibility.}
\label{fig:2d_combined}
\end{figure}

Fig.~\ref{fig:2d_combined}(c) demonstrates that RGDD drives the infidelity down to the theoretical lower bound $\mathcal{I}_{\mathrm{lb}}$. With $M/N \approx 0.5$, the optimized infidelity falls within a factor of two of $\mathcal{I}_{\mathrm{lb}}$. With $M \approx N$, the optimized infidelity is indistinguishable from $\mathcal{I}_{\mathrm{lb}}$ at the $10^{-3}$ level for all tested $N$.
In addition, as shown in Fig.~\ref{fig:2d_combined}(d), RGDD helps to efficiently control the phase-space closure error in 2D settings. 
The closure error from the unpenalized optimization in 2D settings is smaller than that in 1D settings. Introducing the penalty into the objective can control this error at a very low level while maintaining the infidelity minimization.
Therefore, refining power-law interactions in 2D ion crystals with multiple global tones is a  promising route in practice.

\section{Conclusion and Outlook}
\label{sec:conclusion}

This work presents a residual-guided drive design for interaction graph engineering using multiple global bichromatic tones in trapped-ion quantum simulators.
Through multiple numerical examples, the algorithm demonstrates its efficiency in synthesizing a wide variety of Ising interactions with a small number of drive tones. 
RGDD operates within the $(N-1)$-dimensional mode-space span and iteratively adds new tones based on the current residual of the interaction synthesis, placing them near the mode with the largest contribution, followed by a rapid refinement of the parameters for all activated tones.

The numerical trials across long-range and short-range interactions confirm that RGDD can approximate representative interactions with a tone count that scales almost linearly with system size, thereby enhancing the experimental feasibility of the solutions. RGDD reaches the theoretical lower bound on infidelity considerably faster than fixed-frequency protocols. The method also applies directly to 2D ion crystals, where the RGDD results consistently approach the bound in the power-law examples. 
The amplitudes of the optimized drives for all tested interactions span only two orders of magnitude, which is compatible with the tunability of available modulators in experiments. We also note that the total laser power required grows polynomially with system size, as shown in Appendix~\ref{app:omega_distribution}.
To counter diabatic motional excitation, in addition to imposing a lower limit on the detunings, the model introduces an optional penalty on the phase-space closure error in the objective, which effectively reduces the total closure error by orders of magnitude and maintains it at a low level for the tested problem sizes. Collectively, these results make experimental demonstrations on near-term trapped-ion platforms realistic.

The residual-guided optimization may extend to broader interaction engineering in trapped ions and beyond. The mode-space perspective in this work can be generalized beyond the far-detuned Ising regime, such that the residual-guided strategy could be adapted to Dicke interactions~\cite{safavi2018verification,guo2026quantum,bullock2026quantum} and to faster gates near motional resonances. The residual-guided protocol may extend to other optimization problems in quantum technology that involve mixed integer modeling, including neutral atoms~\cite{Cazals2025,chevallier2024variational}, superconducting circuits~\cite{morvan2022optimizing,zhang2025efficient}, and few-body interactions~\cite{katz2022n,katz2023demonstration} or broader motional degree of freedom~\cite{bond2022effect} in trapped ions.

On the other hand, although the theoretical lower bound on fidelity $\mathcal{I}_{\mathrm{lb}}$ precludes perfect synthesis of arbitrary interactions, when combined with single-body operations~\cite{richerme2025multi}, Floquet engineering~\cite{shapira2025programmable,Bassler2023synthesisof}, or tweezers~\cite{bond2022effect,schwerdt2026optical}, global drive design holds great potential for complex Hamiltonian engineering tasks. Lastly, as indicated in Ref.~\cite{kyprianidis2024interaction}, the ordering of ions in interaction graph engineering can be vital in practical applications, and future development of optimization methods should address this aspect.

%\section*{Author Contributions}
%YZ and ZZ contributed equally to this work.

\ack{The authors gratefully acknowledge Philip Richerme and Jeffrey Larson for their constructive comments. 
ZZ acknowledge the support of Shanghai Pujiang Program grant No. 25PJA110. This work is also supported by Shanghai Qi Zhi Institute.}

%\section*{Data Availability}
%The data supporting this work are openly available. 

\bibliographystyle{iopart-num}
\bibliography{apssamp}

\newpage
\appendix

Appendix~\ref{app:closure_tradeoff} examines the joint role of the closure penalty $\lambda_\alpha$ and the gate time $t_g$. Appendix~\ref{app:edge_weight} explains the weighted optimization for interaction graph engineering, which is used in the short-range interaction synthesis in Sec.~\ref{sec:1d_short range}. Appendix~\ref{app:nnls_convergence} compares RGDD with the nonadiabatic equal-spacing tone settings~\cite{shapira2020theory,kyprianidis2024interaction}, which allow for ion motion decoupling with at least $2N+1$ tones. 
Appendix~\ref{app:2d_mode_analysis} elaborates on the technical details of the 2D ion trap examples.
Appendix~\ref{app:omega_distribution} discusses the laser power required by the RGDD-optimized results.

\section{Closure penalty and gate time}
\label{app:closure_tradeoff}

\begin{figure}[h]
\centering
\includegraphics[width=\textwidth]{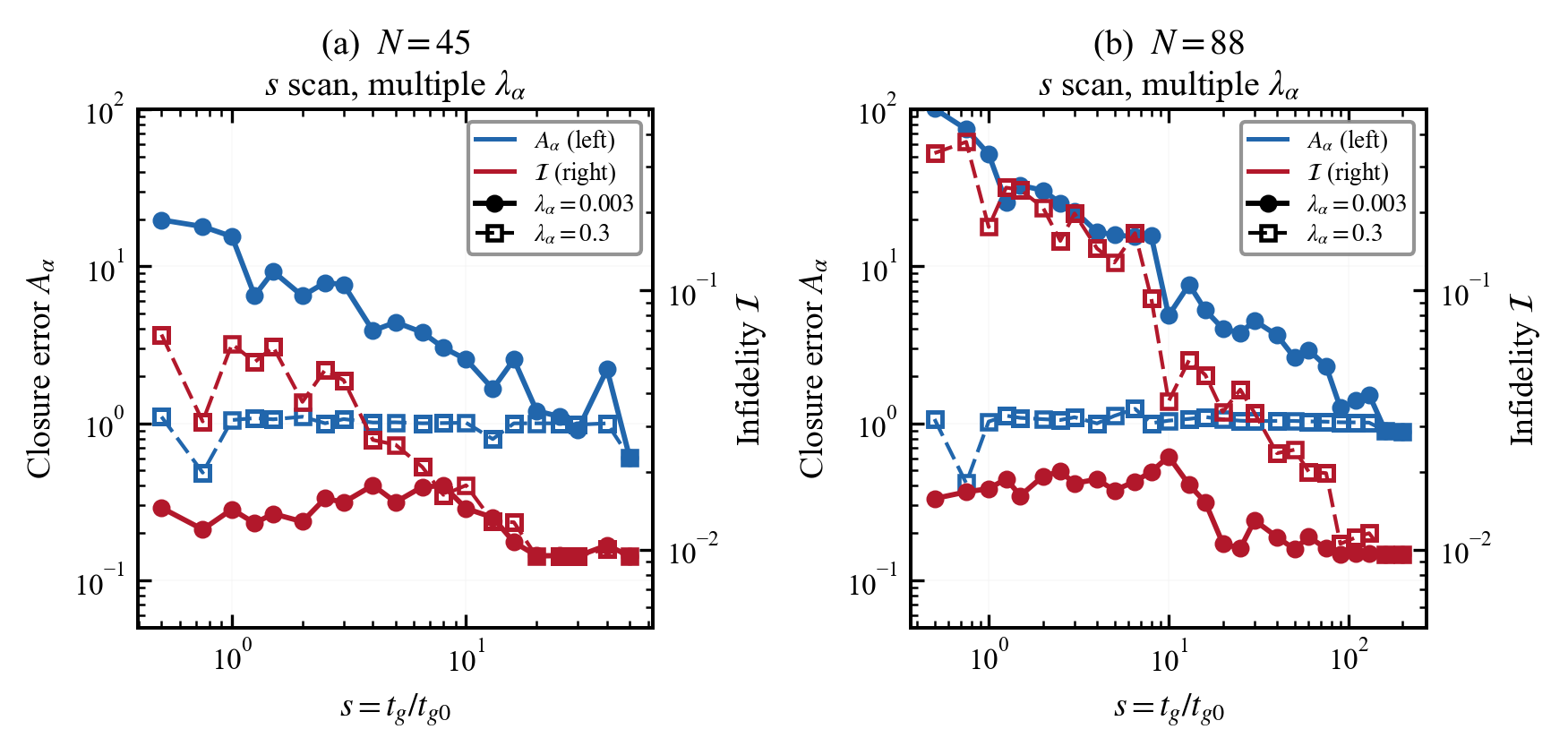}
\caption{Closure error $A_\alpha$ (left axis) and infidelity $\mathcal{I}$ (right axis) versus gate time ratio $s=t_g/t_{g0}$. Circles: $\lambda_\alpha=0.003$ (value used in the main text). Squares: $\lambda_\alpha=0.3$. (a) $N=45$, (b) $N=88$.
}
\label{fig:lambda_tg_tradeoff}
\end{figure}

This appendix examines how the penalty coefficient $\lambda_\alpha$ in the objective~\eqref{eq:closure_objective} and the gate time $t_g$ jointly determine the trade-off between closure error and infidelity. All results shown target on the power-law interaction with $\alpha=1.5$. RGDD optimization is performed toward the infidelity threshold $\mathcal{I}\leq 0.01$ and terminates at the minimum tone count $M_{\min}$ once it is reached. Fig.~\ref{fig:lambda_tg_tradeoff} shows the optimized closure error $A_\alpha$ and infidelity $\mathcal{I}$ across the gate time ratio $s=t_g/t_{g0}$ for $N=45$ and $N=88$, at two penalty values.

Raising the penalty to $\lambda_\alpha=0.3$ immediately confines the closure error to the preset target $A_\text{tar}= 1$. However, at short gate times, overemphasizing this penalty sacrifices fidelity. We also note that the optimization performance saturates beyond a certain gate time. Accordingly, we choose the shortest gate time at which both the closure requirement $A$ and the fidelity threshold $\mathcal{I}$ reach saturated values.

\section{Edge-weight guidance for short-range structured targets}
\label{app:edge_weight}

The cosine infidelity~\eqref{eq:cosine_infid} treats all interaction matrix elements equally, which works well for the long-range interaction synthesis in  Secs.~\ref{sec:1d-powerlaw} and~\ref{sec:2d_prop} of the main text. However, the general decaying nature of interactions among trapped ions calls for a weighted version of the objective for  a short-range target, where the interaction between a pair of distant ions is much weaker than that between neighboring pairs.
The SSH, dimerized, and $J_1$--$J_2$ chains are defined by their nearest-neighbor ($nn$) and next-nearest-neighbor ($nnn$) interactions, while long-range ($lr$) couplings ($|i-j|>2$) are absent from the targets yet 
play an overly large role in the unweighted objective during optimization.  Thus, for short-range interaction optimization, we introduce a weighted inner product
\begin{equation}
\langle \widetilde J^{\rm tar},\mathcal J^{(k)}\rangle_w = \sum_{i<j} w_{ij}\,\widetilde J^{\rm tar}_{ij}\,\mathcal J^{(k)}_{ij},
\qquad w_{ij} = \begin{cases} w_{\rm nn}, & |i-j|=1 \\ w_{\rm nnn}, & |i-j|=2 \\ w_{\rm lr}, & |i-j|>2. \end{cases}
\end{equation}
However, the weights are only incorporated into the calculation of the residual~\eqref{eq:residual}, which ranks the importance of the unselected candidate tones. The calculations of the  objective and all reported infidelities are still based on the unweighted cosine. Therefore, $\mathcal I$ and $\mathcal I_{\rm lb}$ remain comparable across all schemes through this work.

More precisely, in the residual-guided growth of RGDD, at step $q$, the pool for the next tone to add is expanded: besides the $k_q$ modes with the largest unweighted residual coefficients $|c_k^{(q)}|$, we also consider $k_q^{w}$ additional modes with the largest weighted residual coefficients
\begin{equation}
|c_{k,w}^{(q)}|=|\langle R^{(q)},\widetilde{\mathcal J}^{(k)}\rangle_w| .
\end{equation}
We use $k_q=9$ and $k_q^{w}=5$ in this work. The next tone is selected from this pool according to the unweighted infidelity criterion.

\begin{table}[h]
\centering
\caption{Effect of weights on $J_1$--$J_2$, SSH, and Dimer targets at $N=50$ ($k_q=9$, $k_q^{w}=5$). The infidelity threshold is $\mathcal I\le 0.022$ for $J_1$--$J_2$, $\mathcal I\le 0.066$ for SSH, and $\mathcal I\le 0.173$ for dimer ($1.05\,\mathcal I_{\rm lb}$ each). $M_{\min}$ is the smallest tone count reaching the threshold and $\mathcal I$ the infidelity attained there.}
\label{tab:edge_weight_union}
\begin{tabular}{lcc cc cc}
\toprule
& \multicolumn{2}{c}{$J_1$--$J_2$} & \multicolumn{2}{c}{SSH} & \multicolumn{2}{c}{Dimer} \\
\cmidrule(lr){2-3} \cmidrule(lr){4-5} \cmidrule(lr){6-7}
$(w_{\rm nn}, w_{\rm nnn}, w_{\rm lr})$ & $\mathcal I$ & $M_{\min}$ & $\mathcal I$ & $M_{\min}$ & $\mathcal I$ & $M_{\min}$ \\
\midrule
$(1,\,1,\,1)$           & 0.021 & 49 & 0.066 & 34 & 0.171 & 37 \\
$(100,\,1,\,1)$         & 0.021 & 38 & 0.066 & 36 & 0.172 & 34 \\
$(50,\,50,\,1)$         & 0.021 & 38 & 0.066 & 32 & 0.172 & 32 \\
$(1,\,1,\,0.01)$        & 0.021 & 38 & 0.066 & 32 & 0.173 & 30 \\
$\textbf{(25,\,25,\,0.01)}$      & 0.022 & \textbf{36} & 0.067 & \textbf{30} & 0.172 & \textbf{30} \\
\bottomrule
\end{tabular}
\end{table}

Table~\ref{tab:edge_weight_union} compares different weight choices at $N=50$. The unweighted setting $(1,1,1)$ requires $M=49$, $34$, and $37$ for $J_1$--$J_2$, SSH, and dimer, respectively. Suppressing long-range edges, $(1,1,0.01)$, reduces these numbers to $38$, $32$, and $30$, while the settings in this work, $(25,25,0.01)$, gives the lowest tone counts of $36$, $30$, and $30$. At $N=14$, the weighted guidance leaves SSH and dimer unchanged and reduces $J_1$--$J_2$ from $M=11$ to $M=8$. 

We have also tried replacing the unweighted ranking of importance entirely with the weighted version. However, the performance of these trials is worse than that of the combination of both the unweighted and weighted rankings. Therefore, the weights can supplement the rankings, but cannot replace the unweighted importance.

\section{Nonadiabatic protocol baseline}
\label{app:nnls_convergence}

Previous global-drive constructions mainly focus on the far-detuned regime. In this section, we compare RGDD with a simple equal-spacing placement~\cite{shapira2020theory,kyprianidis2024interaction}, which guarantees cancellation of the phonon displacement when a large number of tones are used. The equal tone spacing is fixed as $\Delta\mu=2\pi/t_g$, with the same gate time constrained by the laser power as that used for this work. 
We begin inserting $M$ equal-spacing tones in the vicinity of the lowest zigzag mode, ascending in beatnote frequency. The comparison is performed at $N=14$ for the four short-range targets.

\begin{figure}[htbp]
\centering
\includegraphics[width=0.55\linewidth]{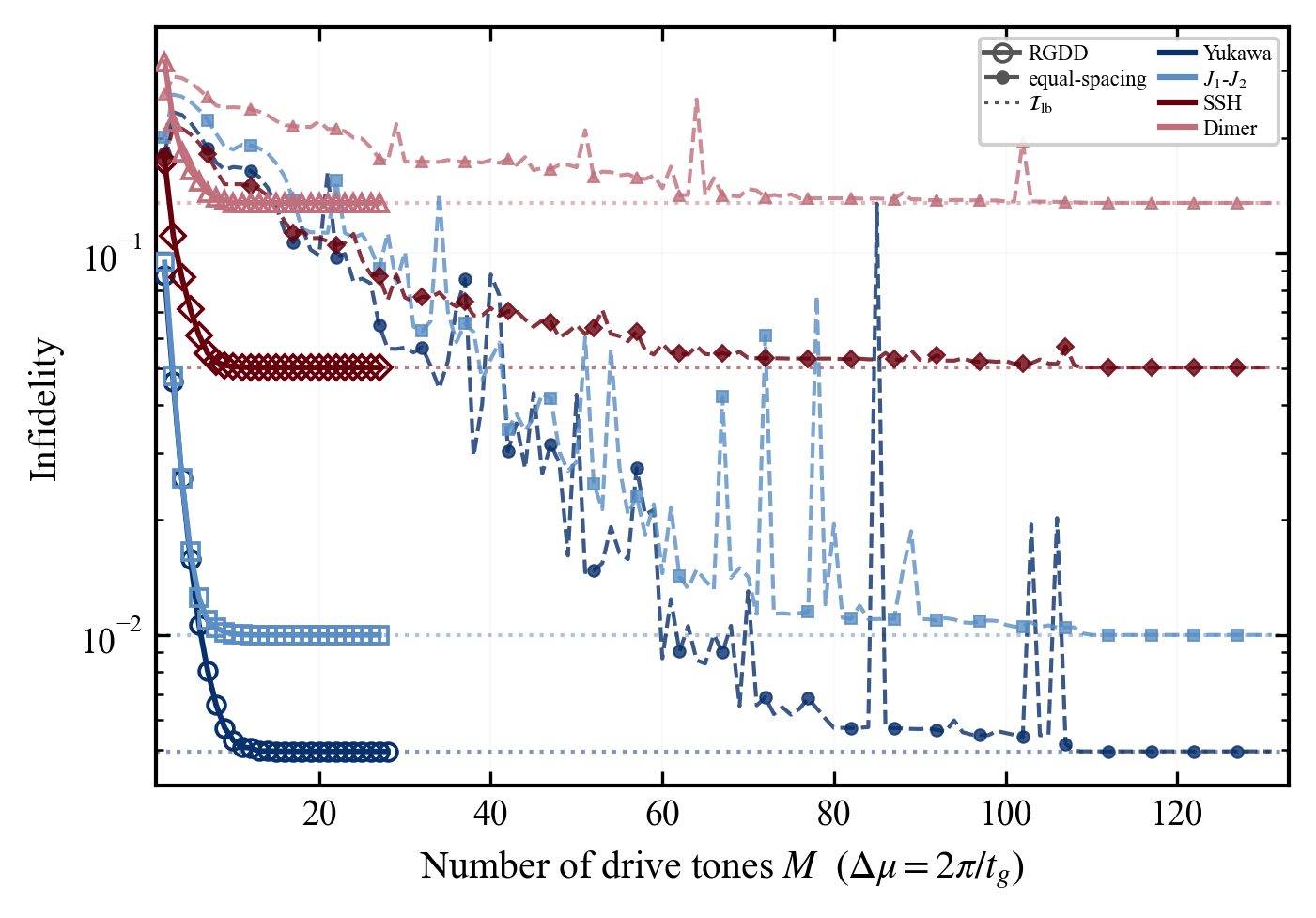}
\caption{Infidelity versus the number of drive tones $M$ at $N=14$.
RGDD is compared with equal-spacing tone placement using the same gate time.
Dotted lines indicate the corresponding lower bounds. These bounds are $\mathcal I_{\rm lb}=0.005$, $0.010$, $0.050$, and $0.135$ for Yukawa, $J_1$--$J_2$, SSH, and dimer, respectively.}
\label{fig:convergence_appendix}
\end{figure}

For the equal-spacing tone protocol, at least $M=2N+1$ tones can be used to decouple the ion motion from the spin interaction~\cite{shapira2020theory}. 
This minimal number of tones required corresponds to their fastest gate time, $t_g=4\pi/\overline{\Delta\omega}$. 
Our default gate time, $t_g=2\pi/(0.1\,\overline{\Delta\omega})$, is five times longer, giving a five times finer tone spacing and requiring about five times more tones to cover the same mode band for the equal-spacing tone placement. The longer gate time reduces the interaction strength required, thereby relaxing the required laser power.
The equal spacing fixes only the beatnote positions, while the tone strengths need to be optimized. Here, for each $M$, the Rabi amplitudes are obtained by bounded least squares to reproduce the target mode coefficients, using the same amplitude bound as RGDD. The small fluctuations in Fig.~\ref{fig:convergence_appendix} arise because this optimization minimizes the absolute fitting error~\cite{shapira2020theory}, whereas the reported infidelity is scale invariant.

Fig.~\ref{fig:convergence_appendix} compares the infidelity as the number of tones increases.
To synthesize the short-range targets with $N=14$, RGDD reaches $\mathcal I\le1.05\,\mathcal I_{\rm lb}$ with only $M=8$--$11$, whereas the equal-spacing placement requires $M=62$--$107$. At sufficiently large $M$, both methods reach the same lower bound $\mathcal I_{\rm lb}$. For the equal-spacing placement, $A_\alpha$ stays around $5$--$65$ for $M\simeq20$--$100$. It falls below unity only at $M=114$--$121$, which verifies the claim in Ref.~\cite{shapira2020theory}. In contrast, RGDD gives $A_\alpha=1.0$--$4.2$ already at $M=8$.  
The closure can be further reduced by including the closure penalty in the optimization.

\section{Supplementary analysis for 2D ion crystals}
\label{app:2d_mode_analysis}

\begin{figure}[htbp]
\centering
\includegraphics[width=\columnwidth]{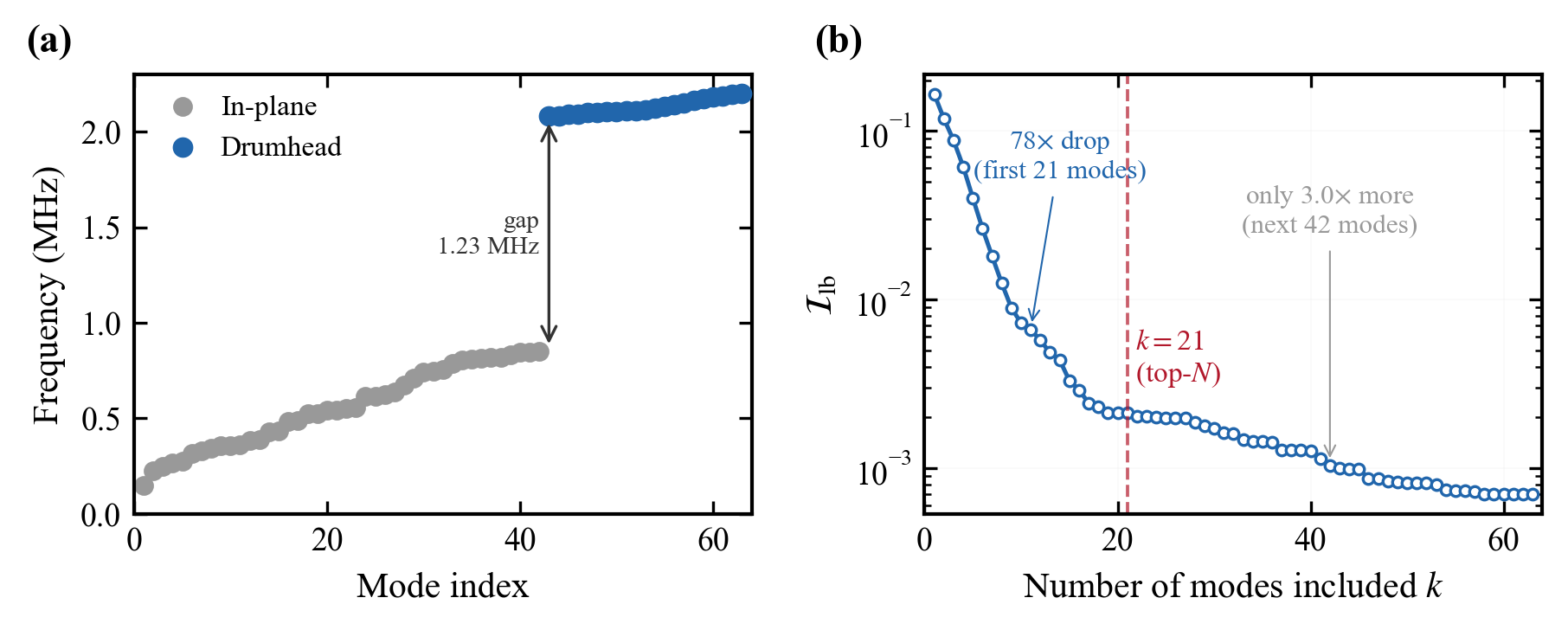}
\caption{(a)~Vibrational mode spectrum of a $N = 21$ zigzag ion crystal. Blue markers: $N$ drumhead modes. Gray markers: $2N$ in-plane modes. The  arrow indicates the spectral gap separating the two sets. (b)~Infidelity lower bound $\mathcal{I}_{\mathrm{lb}}$ versus the number of modes $k$ included in the model, starting from the highest-frequency mode. The red dashed line marks $k = N$, where all drumhead modes are included in the optimization.}
\label{fig:2d_mode_analysis}
\end{figure}

A 2D crystal of $N$ ions supports $3N$ distinct vibrational modes. After removing the three translational zero modes, the remaining $3N - 3$ modes define the accessible $\tilde{J}$-space. Figure~\ref{fig:2d_mode_analysis}(a) displays the mode spectrum of a representative zigzag crystal $N = 21$ with trap frequencies $\{\omega_x, \omega_y, \omega_z\} = 2\pi\times\{0.147, 0.623, 2.200\}\;\mathrm{MHz}$. The spectrum exhibits a large gap between the drumhead and in-plane modes, which suggests the simplification of focusing solely on the former set.

Incorporating additional in-plane modes into the optimization has only a minor impact on the final results, as shown in  Fig.~\ref{fig:2d_mode_analysis}(b). The figure tracks  the theoretical lower bound $\mathcal{I}_{\mathrm{lb}}$ for the power-law interaction with $\alpha=1.5$, which is the same as that in the main text, as a function of the number of modes incorporated into the model, ordered by decreasing frequency. Adding the first $N = 21$ drumhead modes drives $\mathcal{I}_{\mathrm{lb}}$ to $2.1\times10^{-3}$. The subsequent $2N = 42$ in-plane modes improve the bound only by a factor of 3, reaching $\mathcal{I}_{\mathrm{lb}} = 7.0\times10^{-4}$ at the full $3N$ complement. This observation justifies restricting to the top-$N$ drumhead subset throughout this work. Nevertheless, in future studies, the inclusion of additional in-plane modes may offer further opportunities for fine-tuning the results. 

\section{Structure of the RGDD drive amplitudes}
\label{app:omega_distribution}

\begin{figure}[h]
\centering
\includegraphics[width=0.75\linewidth]{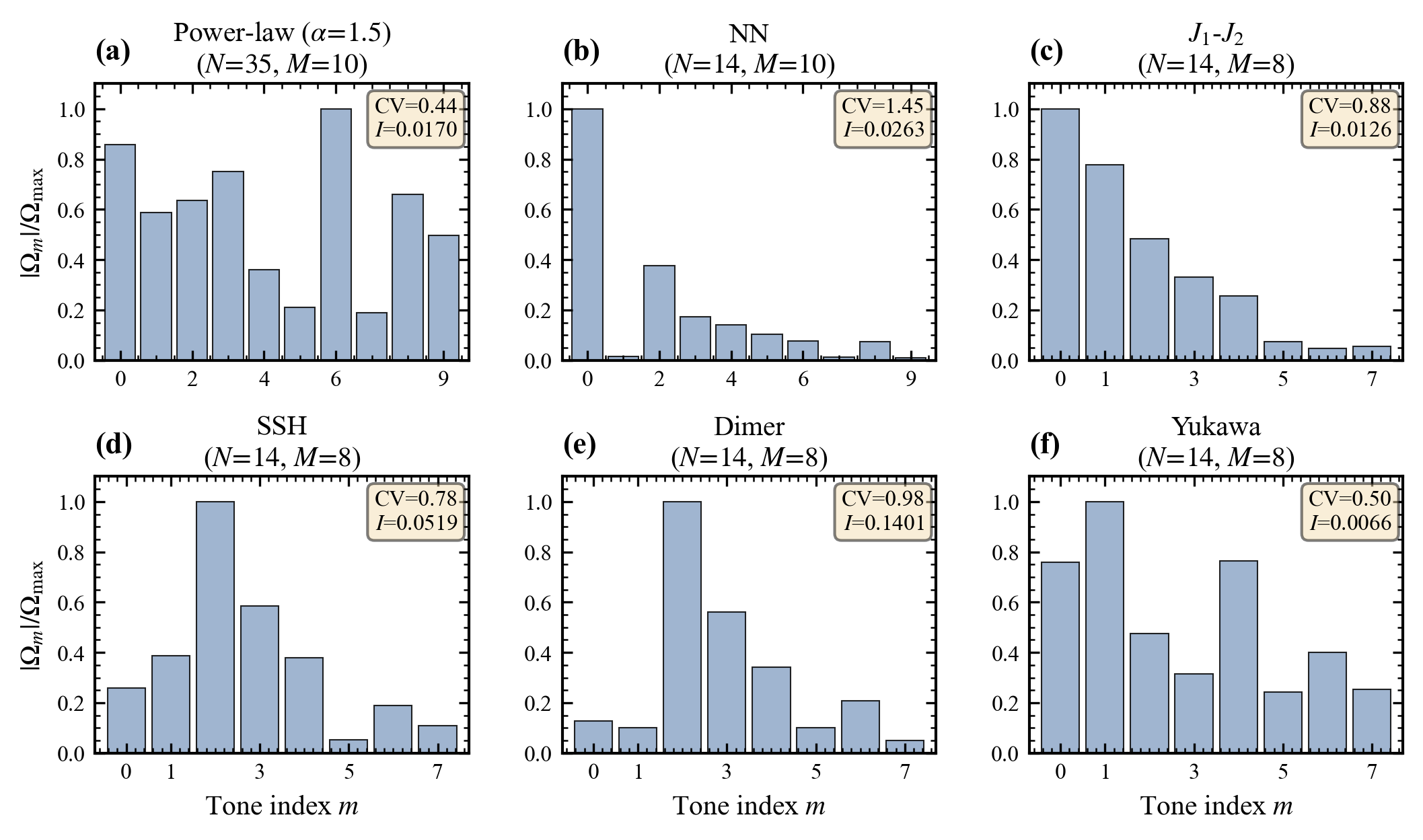}
\caption{Rabi frequencies of the drives designed by RGDD for six target interaction types. Each panel shows $|\Omega_m|/\Omega_{\max}$ for all selected $M$ tones.
CV is the 
variation $\sigma_{|\Omega|}/\overline{|\Omega|}$ of the displayed amplitudes.
Parameters: $N=35$, $M=10$ for the power-law interactions; $N=14$, $M=8$--$10$ for the rest targets. }
\label{fig:omega_bar_by_target}
\end{figure}

This appendix examines the structure of the Rabi frequencies $\{\Omega_m\}_{m=1}^{M}$ for the tones derived from RGDD, covering all six target interaction types studied in the main text. For each target, Fig.~\ref{fig:omega_bar_by_target} displays the normalized amplitudes $|\Omega_m|/\Omega_{\max}$ at the minimal tone count $M_\text{min}$ at which the cosine infidelity $\mathcal{I}$ meets a preset threshold, with all $M_\text{min}$ tones shown in order of frequency.

After RGDD optimization, the resulting Rabi frequencies exhibit no extreme values and are well balanced, with most magnitudes  within the same order of magnitude. This feature is particularly important for future implementations, as laser power and stability can sometimes become a bottleneck in experimental setups. Specifically, the optimized Rabi frequencies for the Yukawa and power-law interactions exhibit smooth distributions, whereas those for the remaining interaction types may contain a few weak tones.
Even in this worst case, the required dynamic range stays within two orders of magnitude, which helps keep the experimental implementation on waveform generators stable.

\begin{figure}[h]
\centering
\includegraphics[width=0.55\linewidth]{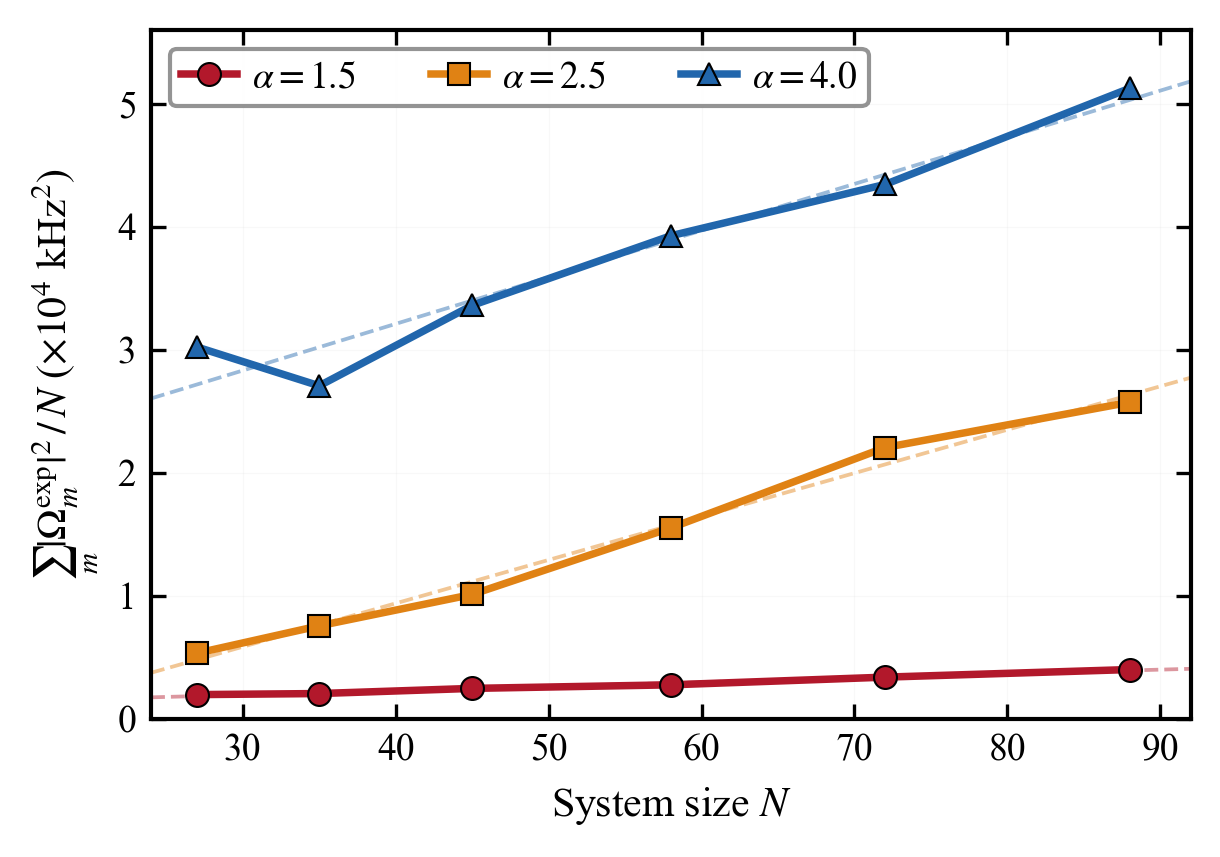}
\caption{Scaling of total drive power for power-law interaction targets. The estimate of total power per ion, $\sum_m|\Omega_m^{\mathrm{exp}}|^2/N$, is plotted as a function of system size $N$ for $\alpha=1.5$, $2.5$, and $4.0$, at the minimal $M_\text{min}$ required to achieve $\mathcal{I}\le 0.05$. The  interactions are Kac-normalized. Dashed lines are linear fits.}
\label{fig:omega_sum_vs_N}
\end{figure}

The total laser power required for the RGDD-optimized parameters grows polynomially with system size, which is an acceptable scaling for experimental feasibility. Fig.~\ref{fig:omega_sum_vs_N} shows the estimate of the average power $\sum_m|\Omega_m^{\mathrm{exp}}|^2/N$ for power-law targets at three exponents $\alpha$ used in Sec.~\ref{sec:1d-powerlaw}, using the minimal number of tones required to reach $\mathcal{I}\le 0.05$. To keep the results consistent with the experimental implementation, the interaction is normalized by a Kac factor~\cite{li2023probing,homrighausen2017anomalous}.
The linear fits in Fig.~\ref{fig:omega_sum_vs_N} give slopes of $0.341$ for $\alpha=1.5$, $3.529$ for $\alpha=2.5$, and $3.793$ for $\alpha=4.0$, in units of $10^{6}~\mathrm{kHz}^2$ per ion.

%\bibliographystyle{iopart-num}

\end{document}